\documentclass[conference]{IEEEtran}
\pdfoutput=1 

\usepackage[nosort]{cite}        
\usepackage{xspace}
\usepackage{amssymb}
\usepackage{url} 
\usepackage[intlimits]{amsmath}  
\usepackage{amsfonts}
\usepackage{bbm}
\usepackage[below]{placeins}

\usepackage[latin1]{inputenc}
  \usepackage[pdftex]{graphicx}

\makeatletter
\def\@IEEEinterspaceratioM{0.265}
\def\@IEEEinterspaceMINratioM{0.1651}
\def\@IEEEinterspaceMAXratioM{0.38}

\def\@IEEEinterspaceratioB{0.31}
\def\@IEEEinterspaceMINratioB{0.19}
\def\@IEEEinterspaceMAXratioB{0.38}
\@IEEEtunefonts
\makeatother

\usepackage{amsfonts}
\usepackage{bbm}
\usepackage{mathrsfs}
\usepackage{xspace}
\usepackage{bm}

\newenvironment{textbmatrix}{	\setlength{\arraycolsep}{2.5pt}%
								\big[\begin{matrix}}{\end{matrix}\big]%
								\raisebox{0.08ex}{\vphantom{M}}}

\def\be{\begin{equation}}
\def\ee{\end{equation}}
\def\een{\nonumber \end{equation}}
\def\mat{\begin{bmatrix}}
\def\emat{\end{bmatrix}}
\def\btm{\begin{textbmatrix}}
\def\etm{\end{textbmatrix}}

\def\ba#1\ea{\begin{align}#1\end{align}}
\def\bs#1\es{\begin{split}#1\end{split}} 
\def\bg#1\eg{\begin{gather}#1\end{gather}} 
\def\bi#1\ei{\begin{itemize}#1\end{itemize}}

\newcommand{\safemath}[2]{\newcommand{#1}{\ensuremath{#2}\xspace}}

\newcommand{\lefto}{\mathopen{}\left}

\DeclareMathOperator{\diag}{diag}			% diagonal matrix
\DeclareMathOperator*{\argmax}{arg\;max}		% arg max
\DeclareMathOperator{\Exop}{\mathbb{E}}		% expectation operator
\newcommand{\Ex}[1]{\ensuremath{\Exop\lefto[#1\right]}} 	% expectation
\safemath{\normal}{\mathcal{N}}				% normal distribution
\safemath{\complexnormal}{\mathcal{CN}}				% complex normal distribution

\safemath{\circnorm}{\mathcal{CN}}			% circ. symm. normal
\safemath{\interior}{\mathrm{Int}}			% interior of a set
\safemath{\dfn}{:=}							% definition
\safemath{\markov}{\leftrightarrow}			% Markov Chain
\safemath{\SNR}{\text{\sc snr}} 				% signal to noise ratio
\safemath{\No}{N_0}							% noise spectral density
\safemath{\Eb}{E_b}							% energy per bit
\safemath{\EbNo}{\frac{\Eb}{\No}}
\safemath{\EsNo}{\frac{\Es}{\No}}

\DeclareMathOperator{\CHop}{\ensuremath{\mathbb{H}}} % channel operator
\safemath{\tvir}{h_{\CHop}}					% time-varying impulse response
\safemath{\tvtf}{L_{\CHop}}					% 	transfer function
\safemath{\spf}{S_{\CHop}}						
\safemath{\bff}{H_{\CHop}}					%	function

\safemath{\ircf}{R_{h}}						% impulse response correlation fn.
\safemath{\scf}{R_{S}}						% scattering function
	\safemath{\CH}{C_{\CHop}}
\safemath{\tvcf}{R_{L}}						% time-varying correlation fn.
\safemath{\bfcf}{R_{H}}						% bi-frequency correlation fn.
\safemath{\RH}{R_{\CHop}}

\safemath{\mi}{I}							% muttal information
\safemath{\capacity}{C}						% capacity

\safemath{\dB}{\,\mathrm{dB}}
\safemath{\dBm}{\,\mathrm{dBm}}
\safemath{\Hz}{\,\mathrm{Hz}}
\safemath{\kHz}{\,\mathrm{kHz}}
\safemath{\MHz}{\,\mathrm{MHz}}
\safemath{\GHz}{\,\mathrm{GHz}}
\safemath{\s}{\,\mathrm{s}}
\safemath{\ms}{\,\mathrm{ms}}
\safemath{\mus}{\,\mathrm{\mu s}}
\safemath{\ns}{\,\mathrm{ns}}
\safemath{\meter}{\,\mathrm{m}}
\safemath{\mm}{\,\mathrm{mm}}
\safemath{\cm}{\,\mathrm{cm}}
\safemath{\km}{\, \mathrm{km}}
\safemath{\m}{\,\mathrm{m}}
\safemath{\J}{\,\mathrm{J}}
\safemath{\K}{\,\mathrm{K}}
\safemath{\bit}{\,\mathrm{bit}}

\safemath{\define}{\triangleq}			% definition

\safemath{\equivalent}{\sim}
\safemath{\distas}{\sim}					% distributed according to

\safemath{\reals}{\mathbb{R}}
\safemath{\positivereals}{\mathbb{R}^{+}}
\safemath{\integers}{\mathbb{Z}}
\safemath{\posint}{\mathbb{Z}_{+}}
\safemath{\naturals}{\mathbb{N}}
\safemath{\complexset}{\mathbb{C}}

\safemath{\setA}{\mathcal{A}}
\safemath{\setB}{\mathcal{B}}
\safemath{\setC}{\mathcal{C}}
\safemath{\setD}{\mathcal{D}}
\safemath{\setE}{\mathcal{E}}
\safemath{\setF}{\mathcal{F}}
\safemath{\setG}{\mathcal{G}}
\safemath{\setH}{\mathcal{H}}
\safemath{\setI}{\mathcal{I}}
\safemath{\setJ}{\mathcal{J}}
\safemath{\setK}{\mathcal{K}}
\safemath{\setL}{\mathcal{L}}
\safemath{\setM}{\mathcal{M}}
\safemath{\setN}{\mathcal{N}}
\safemath{\setO}{\mathcal{O}}
\safemath{\setP}{\mathcal{P}}
\safemath{\setQ}{\mathcal{Q}}
\safemath{\setR}{\mathcal{R}}
\safemath{\setS}{\mathcal{S}}
\safemath{\setT}{\mathcal{T}}
\safemath{\setU}{\mathcal{U}}
\safemath{\setV}{\mathcal{V}}
\safemath{\setW}{\mathcal{W}}
\safemath{\setX}{\mathcal{X}}
\safemath{\setY}{\mathcal{Y}}
\safemath{\setZ}{\mathcal{Z}}
\safemath{\emptySet}{\varnothing}

\safemath{\bma}{\mathbf{a}}
\safemath{\bmb}{\mathbf{b}}
\safemath{\bmc}{\mathbf{c}}
\safemath{\bmd}{\mathbf{d}}
\safemath{\bme}{\mathbf{e}}
\safemath{\bmf}{\mathbf{f}}
\safemath{\bmg}{\mathbf{g}}
\safemath{\bmh}{\mathbf{h}}
\safemath{\bmi}{\mathbf{i}}
\safemath{\bmj}{\mathbf{j}}
\safemath{\bmk}{\mathbf{k}}
\safemath{\bml}{\mathbf{l}}
\safemath{\bmm}{\mathbf{m}}
\safemath{\bmn}{\mathbf{n}}
\safemath{\bmo}{\mathbf{o}}
\safemath{\bmp}{\mathbf{p}}
\safemath{\bmq}{\mathbf{q}}
\safemath{\bmr}{\mathbf{r}}
\safemath{\bms}{\mathbf{s}}
\safemath{\bmt}{\mathbf{t}}
\safemath{\bmu}{\mathbf{u}}
\safemath{\bmv}{\mathbf{v}}
\safemath{\bmx}{\mathbf{x}}
\safemath{\bmy}{\mathbf{y}}
\safemath{\bmz}{\mathbf{z}}

\bmdefine{\biad}{a}
\bmdefine{\bibd}{b}
\bmdefine{\bicd}{c}
\bmdefine{\bidd}{d}
\bmdefine{\bied}{e}
\bmdefine{\bifd}{f}
\bmdefine{\bigd}{g}
\bmdefine{\bihd}{h}
\bmdefine{\biid}{i}
\bmdefine{\bijd}{j}
\bmdefine{\bikd}{k}
\bmdefine{\bild}{l}
\bmdefine{\bimd}{m}
\bmdefine{\bind}{n}
\bmdefine{\biod}{o}
\bmdefine{\bipd}{p}
\bmdefine{\biqd}{q}
\bmdefine{\bird}{r}
\bmdefine{\bisd}{s}
\bmdefine{\bitd}{t}
\bmdefine{\biud}{u}
\bmdefine{\bivd}{v}
\bmdefine{\biwd}{w}
\bmdefine{\bixd}{x}
\bmdefine{\biyd}{y}
\bmdefine{\bizd}{z}

\bmdefine{\bixid}{\xi}
\bmdefine{\bilambdad}{\lambda}
\bmdefine{\bimud}{\mu}
\bmdefine{\bithetad}{\theta}
\bmdefine{\biphid}{\phi}
\bmdefine{\bipi}{\pi}

\safemath{\bmia}{\biad}
\safemath{\bmib}{\bibd}
\safemath{\bmic}{\bicd}
\safemath{\bmid}{\bidd}
\safemath{\bmie}{\bied}
\safemath{\bmif}{\bifd}
\safemath{\bmig}{\bigd}
\safemath{\bmih}{\bihd}
\safemath{\bmii}{\biid}
\safemath{\bmij}{\bijd}
\safemath{\bmik}{\bikd}
\safemath{\bmil}{\bild}
\safemath{\bmim}{\bimd}
\safemath{\bmin}{\bind}
\safemath{\bmio}{\biod}
\safemath{\bmip}{\bipd}
\safemath{\bmiq}{\biqd}
\safemath{\bmir}{\bird}
\safemath{\bmis}{\bisd}
\safemath{\bmit}{\bitd}
\safemath{\bmiu}{\biud}
\safemath{\bmiv}{\bivd}
\safemath{\bmiw}{\biwd}
\safemath{\bmix}{\bixd}
\safemath{\bmiy}{\biyd}
\safemath{\bmiz}{\bizd}

\safemath{\bmxi}{\bixid}
\safemath{\bmlambda}{\bilambdad}
\safemath{\bmmu}{\bimud}
\safemath{\bmtheta}{\bithetad}
\safemath{\bmphi}{\biphid}
\safemath{\bmpi}{\bipi}

\safemath{\bA}{\mathbf{A}}
\safemath{\bB}{\mathbf{B}}
\safemath{\bC}{\mathbf{C}}
\safemath{\bD}{\mathbf{D}}
\safemath{\bE}{\mathbf{E}}
\safemath{\bF}{\mathbf{F}}
\safemath{\bG}{\mathbf{G}}
\safemath{\bH}{\mathbf{H}}
\safemath{\bI}{\mathbf{I}}
\safemath{\bJ}{\mathbf{J}}
\safemath{\bK}{\mathbf{K}}
\safemath{\bL}{\mathbf{L}}
\safemath{\bM}{\mathbf{M}}
\safemath{\bN}{\mathbf{N}}
\safemath{\bO}{\mathbf{O}}
\safemath{\bP}{\mathbf{P}}
\safemath{\bQ}{\mathbf{Q}}
\safemath{\bR}{\mathbf{R}}
\safemath{\bS}{\mathbf{S}}
\safemath{\bT}{\mathbf{T}}
\safemath{\bU}{\mathbf{U}}
\safemath{\bV}{\mathbf{V}}
\safemath{\bW}{\mathbf{W}}
\safemath{\bX}{\mathbf{X}}
\safemath{\bY}{\mathbf{Y}}
\safemath{\bZ}{\mathbf{Z}}

\safemath{\bZero}{\mathbf{0}}

\bmdefine{\biAd}{A}
\bmdefine{\biBd}{B}
\bmdefine{\biCd}{C}
\bmdefine{\biDd}{D}
\bmdefine{\biEd}{E}
\bmdefine{\biFd}{F}
\bmdefine{\biGd}{G}
\bmdefine{\biHd}{H}
\bmdefine{\biId}{I}
\bmdefine{\biJd}{J}
\bmdefine{\biKd}{K}
\bmdefine{\biLd}{L}
\bmdefine{\biMd}{M}
\bmdefine{\biOd}{N}
\bmdefine{\biPd}{O}
\bmdefine{\biQd}{P}
\bmdefine{\biRd}{R}
\bmdefine{\biSd}{S}
\bmdefine{\biTd}{T}
\bmdefine{\biUd}{U}
\bmdefine{\biVd}{V}
\bmdefine{\biWd}{W}
\bmdefine{\biXd}{X}
\bmdefine{\biYd}{Y}
\bmdefine{\biZd}{Z}

\bmdefine{\biDelta}{\Delta}
\bmdefine{\biLambda}{\Lambda}
\bmdefine{\biPhi}{\Phi}
\bmdefine{\biSigma}{\Sigma}
\bmdefine{\biOmega}{\Omega}
\bmdefine{\biTheta}{\Theta}

\safemath{\bimA}{\biAd}
\safemath{\bimB}{\biBd}
\safemath{\bimC}{\biCd}
\safemath{\bimD}{\biDd}
\safemath{\bimE}{\biEd}
\safemath{\bimF}{\biFd}
\safemath{\bimG}{\biGd}
\safemath{\bimH}{\biHd}
\safemath{\bimI}{\biId}
\safemath{\bimJ}{\biJd}
\safemath{\bimK}{\biKd}
\safemath{\bimL}{\biLd}
\safemath{\bimM}{\biMd}
\safemath{\bimN}{\biNd}
\safemath{\bimO}{\biOd}
\safemath{\bimP}{\biPd}
\safemath{\bimQ}{\biQd}
\safemath{\bimR}{\biRd}
\safemath{\bimS}{\biSd}
\safemath{\bimT}{\biTd}
\safemath{\bimU}{\biUd}
\safemath{\bimV}{\biVd}
\safemath{\bimW}{\biWd}
\safemath{\bimX}{\biXd}
\safemath{\bimY}{\biYd}
\safemath{\bimZ}{\biZd}

\safemath{\bDelta}{\bielta}
\safemath{\bLambda}{\biLambda}
\safemath{\bPhi}{\biPhi}
\safemath{\bSigma}{\biSigma}
\safemath{\bOmega}{\biOmega}
\safemath{\bTheta}{\biTheta}

\safemath{\veca}{\bma}
\safemath{\vecb}{\bmb}
\safemath{\vecc}{\bmc}
\safemath{\vecd}{\bmd}
\safemath{\vece}{\bme}
\safemath{\vecf}{\bmf}
\safemath{\vecg}{\bmg}
\safemath{\vech}{\bmh}
\safemath{\veci}{\bmi}
\safemath{\vecj}{\bmj}
\safemath{\veck}{\bmk}
\safemath{\vecl}{\bml}
\safemath{\vecm}{\bmm}
\safemath{\vecn}{\bmn}
\safemath{\veco}{\bmo}
\safemath{\vecp}{\bmp}
\safemath{\vecq}{\bmq}
\safemath{\vecr}{\bmr}
\safemath{\vecs}{\bms}
\safemath{\vect}{\bmt}
\safemath{\vecu}{\bmu}
\safemath{\vecv}{\bmv}
\safemath{\vecw}{\bmw}
\safemath{\vecx}{\bmx}
\safemath{\vecy}{\bmy}
\safemath{\vecz}{\bmz}
\safemath{\vecZero}{\bZero}

\safemath{\vecxi}{\bmxi}
\safemath{\veclambda}{\bmlambda}
\safemath{\vecmu}{\bmmu}
\safemath{\vectheta}{\bmtheta}
\safemath{\vecphi}{\bmphi}
\safemath{\vecpi}{\bmpi}

\safemath{\matA}{\bA}
\safemath{\matB}{\bB}
\safemath{\matC}{\bC}
\safemath{\matD}{\bD}
\safemath{\matE}{\bE}
\safemath{\matF}{\bF}
\safemath{\matG}{\bG}
\safemath{\matH}{\bH}
\safemath{\matI}{\bI}
\safemath{\matJ}{\bJ}
\safemath{\matK}{\bK}
\safemath{\matL}{\bL}
\safemath{\matM}{\bM}
\safemath{\matN}{\bN}
\safemath{\matO}{\bO}
\safemath{\matP}{\bP}
\safemath{\matQ}{\bQ}
\safemath{\matR}{\bR}
\safemath{\matS}{\bS}
\safemath{\matT}{\bT}
\safemath{\matU}{\bU}
\safemath{\matV}{\bV}
\safemath{\matW}{\bW}
\safemath{\matX}{\bX}
\safemath{\matY}{\bY}
\safemath{\matZ}{\bZ}
\safemath{\matZero}{\bZero}

\safemath{\matDelta}{\bDelta}
\safemath{\matLambda}{\bLambda}
\safemath{\matPhi}{\bPhi}
\safemath{\matSigma}{\bSigma}
\safemath{\matOmega}{\bOmega}
\safemath{\matTheta}{\bTheta}

\begin{document}
\IEEEoverridecommandlockouts

\title{Interference Alignment with Limited Feedback}

\author{
\IEEEauthorblockN{Jatin Thukral and Helmut B\"{o}lcskei }
\IEEEauthorblockA{Communication  Technology Laboratory\\
 ETH Zurich, 8092 Zurich, Switzerland\\
E-mail: \{jatin, boelcskei\}@nari.ee.ethz.ch\\} 
\thanks{This work was supported in part by the Swiss National Science Foundation (SNF) under grant No. 200020-109619.}}

\maketitle

\begin{abstract}
We consider single-antenna interference networks where $M$ sources, each with an average transmit power of $P/M$, communicate with $M$ destinations over frequency-selective channels (with $L$ taps each) and each destination has perfect knowledge of its channels from each of the sources. Assuming that there exist error-free non-interfering broadcast feedback links from each destination to all the nodes (i.e., sources and destinations) in the network, we show that naive interference alignment, in conjunction with vector quantization of the impulse response coefficients according to the scheme proposed in Mukkavilli \emph{et al.}, \emph{IEEE Trans. IT}, 2003, achieves full spatial multiplexing gain of $M/2$, provided that the number of feedback bits broadcast by each destination is at least $M(L-1)\log P$. 
\end{abstract}
\IEEEpeerreviewmaketitle

\section{Introduction}
Cadambe and Jafar \cite{Cadambe2008Interference-al} proposed a transmission scheme, called \emph{interference alignment}, for single-antenna interference networks operating over time-selective\footnote{We use the terms time-selective and frequency-selective to denote channels that are selective only in time and only in frequency, respectively.}  channels and showed that this scheme achieves full spatial multiplexing gain. This result depends, however, critically on the assumption of \emph{each} source and \emph{each} destination knowing \emph{all} the channels in the network perfectly. In this paper, we show that  full spatial multiplexing gain is achievable even with partial \emph{channel state information} (CSI) at the sources and the destinations, obtained through limited capacity (error-free) broadcast feedback links.  In particular, we consider an interference network where $M$ single-antenna source-destination pairs, denoted by $\{\mathcal{S}_i,\mathcal{D}_i\}, i=1,\ldots,M,$ communicate concurrently and in the same frequency band over \emph{frequency-selective}  channels with $L$ taps each.\footnote{Interference alignment, as introduced in \cite{Cadambe2008Interference-al}, does not distinguish between time and frequency dimensions. Therefore, although the scheme was originally developed for time-selective channels, it  can equally well be employed for frequency-selective channels. We do not consider time-selective channels as vector quantization of the channel coefficients in such channels would require non-causal feedback.} Each source has an average transmit power of $P/M$ and every destination has perfect knowledge of its channels from each of the sources. Our main contribution is to show that naive interference alignment based on vector-quantized impulse responses, employing the vector quantization scheme proposed for single-user beamforming in \cite{K.K.Mukkavilli2003On-Beamforming-},  achieves full spatial multiplexing gain of $M/2,$ provided that each destination can broadcast at the rate $M(L-1)\log P$ to all the sources and destinations in the network. On a conceptual level, this result shows that rather than aligning interference perfectly by creating completely interference-free signal space dimensions, it suffices to ensure that, as the SNR increases, the interference power in these dimensions remains bounded. 

\paragraph*{Notation}
The superscripts $^T$, $^H$, and $^*$ stand for transposition, Hermitian transpose, and element-wise conjugate, respectively. $\mathcal{CN}(0,\sigma^2)$ denotes a circularly symmetric complex normal distribution with variance $\sigma^2$. Vectors and matrices are set in lower-case and upper-case bold-face letters, respectively. $||\mathbf{x}||$ is the Euclidean norm of the complex vector $\mathbf{x}$ and $|x|$ is the absolute value of the complex scalar $x$. $\Ex{\cdot}$ denotes the expectation operator. $\mathbb{C}^{N\times M}$ is the set of complex matrices with $N$ rows and $M$ columns. The inner product of two column vectors $\mathbf{a}$ and $\mathbf{b}$ of equal dimension is $\mathbf{a}^H\mathbf{b}$. The diagonal matrix of size $N\times N,$ with diagonal entries $a_1,a_2,\ldots,a_N,$ is denoted by $\diag\{a_1,a_2,\ldots,a_N\}$. Square brackets $[\cdot ]$ and circular brackets $(\cdot )$ are used to designate discrete-time and discrete-frequency index, respectively. $\mathbf{A}\circ \mathbf{B}$ is the Hadamard (or element-wise) product of the matrices $\mathbf{A}$ and $\mathbf{B}$. $\log(\cdot)$ stands for logarithm to the base 2 and $j=\sqrt{-1}$. The \emph{discrete Fourier transform} (DFT) of the $N$-point sequence $a[n], n=0,\ldots,N-1,$ is defined as $\mathcal{F}_r\{a[n]\}\triangleq (1/\sqrt{N})\sum_{n=0}^{N-1}a[n]e^{-j2\pi r \frac{n}{N}}$.

%==================================================================

\section{System Model}
\label{sec:MultiAntennaSystemModelSymmetric}
The $L$-tap impulse response of the frequency-selective single-input single-output (SISO) channel between source $\mathcal{S}_k,k=1,\ldots,M,$ and  destination $\mathcal{D}_i, i=1,\ldots,M,$ is given by $h_{i,k}[l], l=0,\ldots,L-1$. The channel coefficients $h_{i,k}[l]$ remain constant throughout the time interval of interest (outage setting) and are drawn independently (across $i,k,l$) from a single continuous probability density function such that $0<|h_{i,k}[l]|<\infty, \forall i,k,l,$ with probability $1$. We use a cyclic signal model (such as in \emph{orthogonal frequency division multiplexing}) to convert the frequency-selective channel $\mathcal{S}_k\rightarrow\mathcal{D}_i,\forall k, i,$ into  $N$ (with $N\gg L$) parallel frequency-flat channels with coefficient $h_{i,k}(r), r=0,\ldots,N-1,$ for the $r$-th tone.  The input-output relation between $\mathcal{S}_i$ and $\mathcal{D}_i$, for the $r$-th tone, is then given by\footnote{We conjugate the channel coefficients for notational simplicity later on.} 
\ba
\label{SISOSymmetricNetworkBasicIORelationFirst}
y_{i}(r) =  h_{i,i}^*(r)x_i(r) + \underbrace{\mathop{\sum_{k\neq i} h_{i,k}^*(r)x_k(r)}}_{\textrm{interference}} +\ z_i(r)
\ea
where $y_i(r)$ is the symbol received at destination $\mathcal{D}_i$, $x_k(r)$ denotes the transmit symbol for source $\mathcal{S}_k$ and $z_i(r)$ is $\mathcal{CN}(0,N_o)$ noise at $\mathcal{D}_i$, all for the $r$-th tone. Defining
\ba
\mathbf{\bar{y}}_i&\triangleq [y_i(0)\ y_i(1)\ \ldots\ y_i(N-1)]^T \notag \\
\mathbf{\bar{x}}_i&\triangleq [x_i(0)\  x_i(1)\ \ldots\ x_i(N-1)]^T \notag\\
\mathbf{\bar{z}}_i&\triangleq [z_i(0)\ z_i(1)\ \ldots\ z_i(N-1)]^T  \notag \\
\mathbf{\bar{H}}_{i,k}&\triangleq \diag\{h_{i,k}(0),h_{i,k}(1),\ldots,h_{i,k}(N-1)\}\notag
\ea
the input-output relation \eqref{SISOSymmetricNetworkBasicIORelationFirst} can be rewritten as
\ba
\mathbf{\bar{y}}_i&=\mathbf{\bar{H}}_{i,i}^H \mathbf{\bar{x}}_i + \sum_{k\neq i}\mathbf{\bar{H}}_{i,k}^H \mathbf{\bar{x}}_k +\mathbf{\bar{z}}_i.
\ea
The transmit signals obey the power constraints 
\ba
\label{SISOSymmetricSystemModelPowerConstraints}
\Ex{|x_k(r)|^2}\leq \frac{P}{M},\  k=1,\ldots,M,\ r=0,\ldots,N-1.
\ea
Finally, we shall also need the channel vector $\mathbf{h}_{i,k}$ and the normalized channel vector $\mathbf{w}_{i,k}$ corresponding to the link $\mathcal{S}_k\rightarrow\mathcal{D}_i,$ defined as $\mathbf{h}_{i,k}=[h_{i,k}[0]\ h_{i,k}[1]\ \cdots\ h_{i,k}[L-1]]^T\in\mathbb{C}^{L\times 1}$ and $\mathbf{w}_{i,k}=\mathbf{h}_{i,k}/\|\mathbf{h}_{i,k}\|\in\mathbb{C}^{L\times 1}$, respectively. 

We assume that each destination $\mathcal{D}_i$ knows its channels from each of the sources $\mathcal{S}_k$ perfectly, that is, $\mathcal{D}_i$ knows $\mathbf{h}_{i,k},\forall k$. There exist dedicated non-interfering error-free broadcast feedback links from each destination $\mathcal{D}_i$ to all the other terminals in the network, that is, to the sources $\mathcal{S}_k,\forall k,$ and to the destinations $\mathcal{D}_k, k\neq i$. In the remainder of the paper, we distinguish between a channel feedback phase during which $N_f$ bits of feedback are broadcast by each destination and a data transmission phase following the channel feedback phase. The channels, being deterministic, are fed back only once during the entire time interval of interest so that the transmission rate loss due to the channel feedback phase can be assumed to be negligible. Denoting the rate of communication for the source-destination pair $\mathcal{S}_i\rightarrow\mathcal{D}_i$ by $R_i,$ and letting $R_{\mathrm{sum}}=\sum_{i=1}^MR_i$, we say that full spatial multiplexing gain is achieved if 
\ba
\label{SISOSymmetricDefinitionFullMuxGain}
\lim_{P\rightarrow\infty} \frac{R_{\mathrm{sum}}}{\log P} = \frac{M}{2}.
\ea
Recall that the spatial multiplexing gain in $M$ source-destination pair single-antenna interference networks is upper-bounded by $M/2$ (see \cite[Th. 1]{Cadambe2008Interference-al}).

%=============================================================
%===================New Section================================
%=============================================================
\section{Interference alignment with perfect CSI at all nodes}
We next briefly review the concept of interference alignment (IA) by adapting the main results of \cite{Cadambe2008Interference-al} to our setup. 

Each source and each destination knows all the channels in the network perfectly. Each source $\mathcal{S}_k$ transmits a linear combination of $d_k$ scalar symbols, $x_k^1,x_k^2,\ldots,x_k^{d_k},$ in $N$ frequency slots by modulating the symbols  onto the transmit direction vectors $\mathbf{v}_k^1,\mathbf{v}_k^2,\ldots,\mathbf{v}_k^{d_k},$ that is,
\ba
\mathbf{\bar{x}}_k&=\sum_{m=1}^{d_k} \mathbf{v}_k^m x_k^m,\qquad k=1,\ldots,M
\ea
where $x_k^m\in\mathbb{C}$, $\mathbf{v}_k^m\in\mathbb{C}^{N\times 1}$ with $\|\mathbf{v}_k^m\|^2=1,$ and $\Ex{|x_k^m|^2}= P/(Md_k),\forall k,m$. Setting $Q=(M-1)(M-2)-1$, the number of data symbols $d_k$ (corresponding to $\mathcal{S}_k$) and the number of tones $N$ are chosen according to (see \cite[Appendix~III]{Cadambe2008Interference-al})
\ba
\label{SymmetricInterferenceSISOFreqSelChoosedk}
d_k&=\begin{cases} 
(t+1)^Q, 	& k=1 \\
t^Q, 		&  k=2,3,\ldots,M 
\end{cases}\\
\label{SymmetricInterferenceSISOFreqSelChooseN}
N&=(t+1)^Q + t^Q
\ea
where $t$ is an auxiliary variable\footnote{We employ the auxiliary variable $t$, partly to simplify our exposition, and partly to keep our presentation consistent with \cite{Cadambe2008Interference-al}. The precise role of $t$ will become clear later.} and the choice of $\mathcal{S}_1$ to transmit $(t+1)^Q$ symbols in $N$ frequency slots, in contrast to $t^Q$ symbols for the other sources, is without loss of generality. Each destination $\mathcal{D}_i$ computes the projections of its received signal $\mathbf{\bar{y}}_i$ onto $d_i$ receive direction vectors $\mathbf{u}_i^1,\mathbf{u}_i^2,\ldots,\mathbf{u}_i^{d_i}$ resulting in a total of $\sum_{i=1}^M d_i$ effective input-output relations given by  
\ba
(\mathbf{u}_i^m)^H\mathbf{\bar{y}}_i &=(\mathbf{u}_i^m)^H\mathbf{\bar{H}}_{i,i}^H \mathbf{v}_i^{m} x_i^{m} +\underbrace{\sum_{ p\neq m} (\mathbf{u}_i^m)^H\mathbf{\bar{H}}_{i,i}^H \mathbf{v}_i^{p} x_i^{p}}_{\text{interference}}\notag\\
\label{IAPerfectCSIEffectiveIO}
& \quad + \underbrace{\sum_{k\neq i}\sum_{p=1}^{d_k}(\mathbf{u}_i^m)^H\mathbf{\bar{H}}_{i,k}^H \mathbf{v}_k^{p} x_k^{p}}_{\text{interference}}  +(\mathbf{u}_i^m)^H\bar{\mathbf{z}}_i
\ea
for $m=1,\ldots,d_i, i=1,\ldots,M,$ where $\mathbf{u}_i^m\in\mathbb{C}^{N\times 1}$ with $\|\mathbf{u}_i^m\|^2=1,\forall i,m$. Choosing $x_k^m,\forall k,m,$ to be i.i.d. Gaussian, treating the two interference terms in \eqref{IAPerfectCSIEffectiveIO} as additional noise, and assuming that $\mathcal{D}_i$ knows the effective channel  coefficient $(\mathbf{u}_i^m)^H\mathbf{\bar{H}}_{i,i}^H\mathbf{v}_i^m,\forall m,$ perfectly, the rate of communication over the link $\mathcal{S}_i\rightarrow\mathcal{D}_i$ is lower-bounded according to 
\ba
\label{IAReviewPerfectCSIFirstRateExpr}
R_i&\geq 
\frac{1}{N}\sum_{m=1}^{d_i} \log\Bigg( 1+ \frac{\frac{P}{Md_i}|(\mathbf{u}_i^m)^H\mathbf{\bar{H}}_{i,i}^H\mathbf{v}_i^m|^2}{ \mathcal{I}_{i,1} + \mathcal{I}_{i,2}  + N_o}    \Bigg)
\ea
with\\[-13mm]
\ba
\mathcal{I}_{i,1}&=\underset{p\neq m}{\sum}\frac{P}{Md_i}\big|(\mathbf{u}_i^m)^H\mathbf{\bar{H}}_{i,i}^H \mathbf{v}_i^{p}\big|^2 \\
\mathcal{I}_{i,2}&=\underset{k\neq i}{\sum} \sum_{p =1}^{d_k}\frac{P}{Md_k}\big|  (\mathbf{u}_i^m)^H\mathbf{\bar{H}}_{i,k}^H \mathbf{v}_k^{p}\big|^2.
\ea

Each source $\mathcal{S}_k$ computes, based on its channel knowledge, transmit direction vectors $\mathbf{v}_k^m,m=1,\ldots,d_k,$ and each destination $\mathcal{D}_i$ computes, based on its channel knowledge, receive direction vectors $\mathbf{u}_i^m,m=1,\ldots,d_i,$ that together satisfy the following three  sets of conditions:
\ba
\label{SymmetricIABasicCondition3}
|(\mathbf{u}_i^m)^H\mathbf{\bar{H}}_{i,i}^H \mathbf{v}_i^{m}|&\geq c > 0, \quad \ \forall i, m\\
\label{SymmetricIABasicCondition1}
(\mathbf{u}_i^m)^H\mathbf{\bar{H}}_{i,i}^H \mathbf{v}_i^{p}&=0, \qquad \quad \forall i, \forall m\neq p\\
\label{SymmetricIABasicCondition2}
(\mathbf{u}_i^m)^H\mathbf{\bar{H}}_{i,k}^H \mathbf{v}_k^{p}&=0, \qquad \quad \forall k\neq i, \forall m, p
\ea
with the constant $c$ independent of $P$. It then follows that $\mathcal{I}_{i,1}=\mathcal{I}_{i,2}=0,\forall i,$  and the spatial multiplexing gain achieved by IA is lower-bounded according to 
\ba
\lim_{P\rightarrow\infty} \frac{R_{\mathrm{sum}}}{\log P}&\geq \lim_{P\rightarrow\infty}\frac{\overset{M}{\underset{i=1}{\sum}}\ \overset{d_i}{\underset{m=1}{\sum}} \log\bigg( 1+ \frac{\frac{P}{Md_i}|(\mathbf{u}_i^m)^H\mathbf{\bar{H}}_{i,i}^H\mathbf{v}_i^m|^2}{ N_o} \bigg)}{N\log P}\notag\\
&=\frac{\sum_{i=1}^Md_i}{N}= \frac{(t+1)^Q+(M-1)t^Q}{(t+1)^Q + t^Q}\ \stackrel{t\rightarrow\infty}{\longrightarrow} \ \frac{M}{2}\notag
\ea
that is, full spatial multiplexing gain, in the sense of \eqref{SISOSymmetricDefinitionFullMuxGain}, 
is achieved. Under the assumption of every node (i.e., every source and every destination) knowing all the channels in the network  perfectly, one way to find vectors $\mathbf{u}_{i}^{m}, \mathbf{v}_{i}^{p}$ satisfying \eqref{SymmetricIABasicCondition3}-\eqref{SymmetricIABasicCondition2}  is provided in \cite[Appendix~III]{Cadambe2008Interference-al}. The basic idea is  that each $\mathcal{S}_k$ computes, based on its knowledge of all the channels in the network, a set of linearly independent transmit direction vectors $\mathbf{v}_k^1,\mathbf{v}_k^2,\ldots,\mathbf{v}_k^{d_k}$ such that \emph{all} the vectors corresponding to interference from $\mathcal{S}_k,k\neq i,$ at $\mathcal{D}_i$ (that is, the vectors $\mathbf{\bar{H}}_{i,k}^H\mathbf{v}_{k}^p, \forall k\neq i, p=1,\ldots,d_k$) span an $(N-d_i)$-dimensional complex subspace of $\mathbb{C}^N$. Consequently, $d_i$ dimensions remain completely interference-free. Each $\mathcal{D}_i$, in turn, computes, based on its knowledge of all the channels in the network, a set of $d_i$ unit-norm receive direction vectors $\mathbf{u}_i^1,\mathbf{u}_i^2,\ldots,\mathbf{u}_i^{d_i}$  that spans the $d_i$-dimensional interference-free subspace corresponding to the link $\mathcal{S}_i\rightarrow\mathcal{D}_i$, thereby satisfying \eqref{SymmetricIABasicCondition2}. Moreover, it was shown in \cite[Appendix~III]{Cadambe2008Interference-al} that if the vectors $\mathbf{\bar{H}}_{i,i}^H\mathbf{v}_i^m, \forall m,$ along with the vectors $\mathbf{\bar{H}}_{i,k}^H\mathbf{v}_k^p,\forall k\neq i,\forall p,$ span $\mathbb{C}^N,$ then $\mathcal{D}_i$ can choose the $d_i$ receive direction vectors $\mathbf{u}_i^m,\forall m,$  such that along with \eqref{SymmetricIABasicCondition2}, both \eqref{SymmetricIABasicCondition3} and \eqref{SymmetricIABasicCondition1} are satisfied as well. It turns out that, in the frequency-selective case, this is possible provided that $L> ((t+1)^Q-1)/(3tQ)$ (the proof of this statement is similar to \cite[Th. 6.4]{Grokop2008Interference-al} and the details are provided in \cite{Thukral2009Spatial-multipl}). 

The developments in the remainder of this paper are based on the simple observation that if the interference power terms $\mathcal{I}_{i,1}$ and $\mathcal{I}_{i,2},$ for all $i,$ are not equal to zero, but upper-bounded by a constant\footnote{To be precise, full spatial multiplexing gain is achieved even if $\mathcal{I}_{i,1}$ and $\mathcal{I}_{i,2}$ scale as a function of $P,$ say $f(P),$ such that $\lim_{P\rightarrow\infty}\frac{\log f(P)}{\log P} = 0$.  Relegating the details to \cite{Thukral2009Spatial-multipl}, we note, however, that this does not result in a reduction of the required feedback rate scaling (in $P$).} \emph{independent of $P$}, full spatial multiplexing gain is still achieved. The key to realizing this will be a vector quantization scheme, which  satisfies \eqref{SymmetricIABasicCondition3} and ensures that both $|(\mathbf{u}_i^m)^H\mathbf{\bar{H}}_{i,i}^H \mathbf{v}_i^{p}|^2,\forall i,\forall m\neq p, $ and $|(\mathbf{u}_i^m)^H\mathbf{\bar{H}}_{i,k}^H \mathbf{v}_k^{p}|^2,\forall k\neq i,$ $\forall m,p,$  scale as $1/P$ when $P\rightarrow\infty$. It will turn out  that the vector quantization scheme developed in \cite{K.K.Mukkavilli2003On-Beamforming-} and \cite{Love2003Grassmannian-be} for beamforming in single-user frequency-flat multi-input multi-output (MIMO) channels satisfies this condition.

%=============================================================
%===================New Section================================
%=============================================================

\section{Interference alignment with limited feedback}
We start by recalling that each destination $\mathcal{D}_i$ knows the channel coefficient vectors $\mathbf{h}_{i,k},\forall k,$ (and hence, $\mathbf{w}_{i,k},\forall k$) perfectly. Knowledge of $\mathbf{w}_{i,k},\forall k,$ at all the sources and all the other destinations is obtained through feedback. Specifically, each $\mathcal{D}_i$ broadcasts, during the channel feedback phase, quantized versions of $\mathbf{w}_{i,k},\forall k,$ to $\mathcal{S}_k,\forall k,$ and $\mathcal{D}_k,\forall k\neq i$. We shall next describe the vector quantization and feedback scheme used.

\subsection{The vector quantization and feedback scheme}
The vector quantization scheme works on unit norm vectors and quantizes the vector $\mathbf{w}_{i,k}\in\mathbb{C}^{L\times 1}$ to the unit norm vector $\mathbf{\hat{w}}_{i,k}\in\mathbb{C}^{L\times 1}$ using $N_d$ bits. The corresponding quantizer codebook $\mathcal{A}$ therefore contains $2^{N_d}$ vectors, that is, $\mathcal{A}=\{\mathbf{\hat{w}}_1, \mathbf{\hat{w}}_2,\ldots,\mathbf{\hat{w}}_{2^{N_d}}\}$. The quantization policy is as follows:    
\ba
\label{SymmetricSISOFreqSelQuantizationRule}
 \mathbf{\hat{w}}_{i,k}&=\argmax_{\mathbf{\hat{w}}_l\in\mathcal{A}}\left\{ |\mathbf{\hat{w}}_l^H\mathbf{w}_{i,k}|\right\}.
\ea
The quantizer codebook is chosen as the solution of the following \emph{Grassmannian line-packing problem \cite{Conway1996Packing-lines-p}: Find the maximum number of unit-magnitude vectors in $\mathbb{C}^L$ such that the absolute value  of the inner product between any two of the vectors is less than $\cos(\delta)$, where $\delta\in(0,\pi/2]$.} Here, $\delta$ is an auxiliary variable whose role will become clear later.

\emph{The line-packing problem:}  If $N_{\mathrm{pack}}$ is the solution to the line-packing problem and  
$\{\mathbf{p}_1,\mathbf{p}_2,\ldots,$ $\mathbf{p}_{N_{\mathrm{pack}}}\}$
 is a set of vectors corresponding to this solution, we choose
  \ba
 2^{N_d}&=N_{\mathrm{pack}}\qquad \text{and set}\notag  \\
\label{MultiAntennaFlatFadingChoiceOfQuantizationVectors}
\{\mathbf{\hat{w}}_1,\mathbf{\hat{w}}_2,\ldots,\mathbf{\hat{w}}_{2^{N_d}}\}&=\{\mathbf{p}_1,\mathbf{p}_2,\ldots,\mathbf{p}_{N_{\mathrm{pack}}}\}.
\ea
This approach was used in  \cite{K.K.Mukkavilli2003On-Beamforming-} and \cite{Love2003Grassmannian-be} for beamforming in single-user MIMO channels. 

\subsubsection*{Quantization error} We define the quantization error $\Delta_d(\mathbf{w}_{i,k},\mathbf{\hat{w}}_{i,k})$ as 
\ba
	\Delta_d(\mathbf{w}_{i,k},\mathbf{\hat{w}}_{i,k})&\triangleq \sqrt{1-|\mathbf{w}_{i,k}^H\mathbf{\hat{w}}_{i,k}|^2 }.\notag
\ea
The maximum quantization error $\Delta_d^{\mathrm{max}}$ is then given by 
\ba
\label{MultiAntennaFlatFadingMaxErrorDirection}
\Delta_d^{\mathrm{max}}=\max_{\mathbf{x}\in\mathbb{C}^L,\|\mathbf{x}\|=1}\sqrt{1-|\mathbf{x}^H\mathbf{\hat{w}}_\mathbf{x}|^2 }
\ea
where $\mathbf{\hat{w}}_\mathbf{x}\in\mathcal{A}$ is the unit-magnitude quantized version of $\mathbf{x}\in\mathbb{C}^{L}$ obtained according to  \eqref{SymmetricSISOFreqSelQuantizationRule}. 

We will need an upper bound on $\Delta_d^{\mathrm{max}}$ in terms of $N_d$. While such a bound is known \cite{K.K.Mukkavilli2003On-Beamforming-}, we will provide a derivation, partly for completeness, and partly to get the bound in a form required for our proof. We start by noting the following two key properties of the  chosen set of quantization vectors:
 \begin{enumerate}
 \renewcommand{\labelenumi}{\roman{enumi}}
 \item[i)] The following relation holds between $N_d$ and $\sin(\delta)$ (see \cite[Th. 3]{Love2003Grassmannian-be}): \\[-8mm] 
\ba
2^{N_d}&\leq \left(\frac{\sin(\delta)}{2}\right)^{-2(L-1)}\notag\\
\label{MultiAntennaFlatFadingBoundFromLoveEtAl}
\Rightarrow \sin(\delta)&\leq 2\left(\frac{1}{2^{\frac{N_d}{2(L-1)}}}\right).
\ea
 \item[ii)] The maximum quantization error $\Delta_d^{\mathrm{max}}$ is upper-bounded by $\sin(\delta)$. This can be proved by contradiction as follows:
 \begin{itemize}
\item Let us assume that $\Delta_d^{\mathrm{max}} > \sin(\delta)$. 
\item Further, let $\mathbf{x}_o$ be the unit-magnitude vector, quantized to $\mathbf{\hat{w}}_{\mathbf{x}_o}$, that corresponds to the maximum quantization error, that is,
\ba
\mathbf{x}_o&=\argmax_{\mathbf{x}\in\mathbb{C}^L,\|\mathbf{x}\|=1}\sqrt{1-\left|\mathbf{x}^H\mathbf{\hat{w}}_{\mathbf{x}}\right|^2}.
\ea
Then, we have
\ba
\sqrt{1-\left|\mathbf{x}_o^H\mathbf{\hat{w}}_{\mathbf{x}_o}\right|^2} &>\sin(\delta)\\
\Rightarrow \left|\mathbf{x}_o^H\mathbf{\hat{w}}_{\mathbf{x}_o}\right|^2&<1-\sin^2(\delta)=\cos^2(\delta)\\
\Rightarrow \left|\mathbf{x}_o^H\mathbf{\hat{w}}_{\mathbf{x}_o}\right|&<\cos(\delta)\\
\label{MultiAntennaFlatFadingContradictionProofArgumentFlatFading}
\Rightarrow \left|\mathbf{x}_o^H\mathbf{{p}}_{l}\right|&<\cos(\delta),\ l=1,\ldots,2^{N_d}.
\ea
\item However, \eqref{MultiAntennaFlatFadingContradictionProofArgumentFlatFading} implies that in the line-packing problem, there exists a solution set of vectors $\{\mathbf{p}_1,\mathbf{p}_2,\ldots,\mathbf{p}_{2^{N_d}}, \mathbf{x}_o\}$, such that the absolute value of the inner product between any two vectors is less than $\cos(\delta)$. In other words, $2^{N_d}$ (and consequently $N_{pack}$) is not the maximum possible number of unit-magnitude vectors with the absolute value of the inner product between any two vectors being less than $\cos(\delta)$ and can hence not be the solution of the line-packing problem, which results in a contradiction. Thus, our premise $\Delta_d^{\mathrm{max}}> \sin(\delta)$ must be incorrect and we must necessarily have
\ba
\label{MultiAntennaFlatFadingSinHalfErrorBound}
\Delta_d^{\mathrm{max}}&\leq \sin(\delta).
\ea
\end{itemize}
\end{enumerate}
Inserting \eqref{MultiAntennaFlatFadingSinHalfErrorBound} into \eqref{MultiAntennaFlatFadingBoundFromLoveEtAl}, we get the desired upper bound: 
 \ba
 \label{MISOInterferenceFlatFadingFeedbackTempUB}
\Delta_d^{\mathrm{max}} &\leq 2\left(\frac{1}{2^{\frac{N_d}{2(L-1)}}}\right).
 \ea 

\subsubsection*{Number of feedback bits} 
During the channel feedback phase, each destination $\mathcal{D}_i$ broadcasts $N_d$ bits for the realization $\mathbf{\hat{w}}_{i,k}, k=1,\ldots,M,$ to all sources and destinations (except to itself, of course) in the network, resulting in a total of 
\ba
\label{IALimitedFeedbackConvertNdToNf}
N_f=MN_d\quad \text{bits}
\ea
 being broadcast by each destination. Each source $\mathcal{S}_k$ therefore receives a total of $M^2N_d$ bits of (error-free) feedback from all the destinations and each destination $\mathcal{D}_i$ receives a total of $(M-1)MN_d$ bits from the destinations $\mathcal{D}_k,k\neq i.$  Each source and each destination can therefore recreate the quantized normalized channel vectors $\mathbf{\hat{w}}_{i,k},\forall i, k,$ and the key to proving the main result of this paper is to determine a value of $N_d$ that ensures the achievability of full spatial multiplexing gain with naive IA based on $\mathbf{\hat{w}}_{i,k},\forall i, k$.

%============
\subsection{Transmission scheme and achievability of full spatial multiplexing gain}
Each source and each destination first converts its received $L$-dimensional quantized vectors $\mathbf{\hat{w}}_{i,k},\forall i, k,$ into $N$-dimensional vectors $[\hat{w}_{i,k}[0]\ \hat{w}_{i,k}[1]\ \ldots\ $ $\hat{w}_{i,k}[N-1] ]^T$ through zero-padding. It then computes the $N$-point DFTs 
\ba
\hat{w}_{i,k}(r)=\mathcal{F}_r\{\hat{w}_{i,k}[n]\},\qquad r=0,\ldots,N-1
\ea
and organizes the results into the quantized channel matrices  
\ba
\mathbf{ \widehat{W}}_{i,k}&=\diag\{\hat{w}_{i,k}(0),\hat{w}_{i,k}(1),\ldots,\hat{w}_{i,k}(N-1)\}.
\ea 

IA is now performed naively assuming that $\mathbf{\bar{H}}_{i,k}=\mathbf{ \widehat{W}}_{i,k},$ $\forall i,k,$ that is,  each source $\mathcal{S}_k$ computes its transmit direction vectors $\mathbf{\hat{v}}_{k}^m, m=1,\ldots,d_k,$ and each destination $\mathcal{D}_i$ computes its receive direction vectors $\mathbf{\hat{u}}_{i}^m, m=1,\ldots,d_i,$ from $\mathbf{\widehat{W}}_{i,k},\forall i,k$ (rather than from $\mathbf{\bar{H}}_{i,k}$). $\mathcal{S}_k$ transmits a linear combination of $d_k$ scalar symbols, $x_k^1,x_k^2,\ldots,x_k^{d_k},$ in $N$ frequency-slots by modulating the symbols onto the vectors $\mathbf{\hat{v}}_k^1,\mathbf{\hat{v}}_k^2,\ldots,\mathbf{\hat{v}}_k^{d_k},$ that is,\\[-4mm]
\ba
\mathbf{\bar{x}}_k&=\sum_{m=1}^{d_k} \mathbf{\hat{v}}_k^m x_k^m,\qquad k=1,\ldots,M
\ea
where $x_k^m\in\mathbb{C}$, $\mathbf{\hat{v}}_k^m\in\mathbb{C}^{N\times 1}$ with $\|\mathbf{\hat{v}}_k^m\|^2=1,$ and $\Ex{|x_k^m|^2}= P/(Md_k),\forall k,m$. The number of data symbols $d_k$ (corresponding to $\mathcal{S}_k$) and the number of tones $N$ are chosen according to \eqref{SymmetricInterferenceSISOFreqSelChoosedk} and \eqref{SymmetricInterferenceSISOFreqSelChooseN}, respectively. Each destination $\mathcal{D}_i$ computes the projections of its received signal $\mathbf{\bar{y}}_i$ onto the receive direction vectors $\mathbf{\hat{u}}_i^1,\mathbf{\hat{u}}_i^2,\ldots,\mathbf{\hat{u}}_i^{d_i}$ resulting in a total of $\sum_{i=1}^M d_i$ effective input-output relations given by 
\ba
\label{SymmetricInterferenceFreqSelLimitedFeedbackBasicIO}
&(\mathbf{\hat{u}}_i^m)^H\mathbf{\bar{y}}_i=(\mathbf{\hat{u}}_i^m)^H\mathbf{\bar{H}}_{i,i}^H \mathbf{\hat{v}}_i^{m} x_i^{m}
+\sum_{p\neq m} (\mathbf{\hat{u}}_i^m)^H\mathbf{\bar{H}}_{i,i}^H \mathbf{\hat{v}}_i^{p} x_i^{p}  \notag\\
&\qquad+ \sum_{k\neq i}\sum_{p\hspace{0.5mm}=1}^{d_k}(\mathbf{\hat{u}}_i^{m})^H\mathbf{\bar{H}}_{i,k}^H \mathbf{\hat{v}}_k^{p} x_k^{p}  +(\mathbf{\hat{u}}_i^m)^H\bar{\mathbf{z}}_i,\quad \forall i,m
\ea
where $\mathbf{\hat{u}}_i^m\in\mathbb{C}^{N\times 1}$ with $\|\mathbf{\hat{u}}_i^m\|^2=1,\forall i,m$. Defining $\mathbf{\bar{h}}_{i,k} \triangleq [h_{i,k}(0)\ h_{i,k}(1)\ \ldots \ h_{i,k}(N-1)]^T$ and $\mathbf{\hat{b}}_{i,k}^{m,p}   \triangleq  (\mathbf{\hat{u}}_i^m)^*\circ \mathbf{\hat{v}}_k^p,$ we can rewrite the input-output relations \eqref{SymmetricInterferenceFreqSelLimitedFeedbackBasicIO} as
\ba
\label{SymmetricInterferenceFreqSelLimitedFeedbackModifiedIO}
(\mathbf{\hat{u}}_i^m)^H\mathbf{\bar{y}}_i 
&=\mathbf{\bar{h}}_{i,i}^H \mathbf{\hat{b}}_{i,i}^{m,m} x_i^{m} +\sum_{p\neq m} \mathbf{\bar{h}}_{i,i}^H \mathbf{\hat{b}}_{i,i}^{m,p} x_i^{p} \notag\\
&+ \sum_{k\neq i}\sum_{p\hspace{0.5mm}=1}^{d_k}\mathbf{\bar{h}}_{i,k}^H \mathbf{\hat{b}}_{i,k}^{m,p} x_k^{p}  +(\mathbf{\hat{u}}_i^m)^H\bar{\mathbf{z}}_i,\quad \forall i,m.
\ea
Choosing $x_i^m,\forall i,m,$ to be i.i.d. Gaussian, treating the two interference terms in \eqref{SymmetricInterferenceFreqSelLimitedFeedbackModifiedIO} as additional noise and assuming that the destination $\mathcal{D}_i$ knows the effective channel coefficients $\mathbf{\bar{h}}_{i,i}^H\mathbf{\hat{b}}_{i,i}^{m,m},\forall m,$ perfectly, the rate of communication over the link $\mathcal{S}_i\rightarrow\mathcal{D}_i$ is then lower-bounded according to
\ba
\label{IALimitedFeedbackSingleUserRateExpression}
R_i \geq \frac{1}{N}\sum_{m=1}^{d_i} \log\Bigg( 1+ \frac{\frac{P}{Md_i}|\mathbf{\bar{h}}_{i,i}^H\mathbf{\hat{b}}_{i,i}^{m,m}|^2}{ \mathcal{I}_{i,1} +\mathcal{I}_{i,2}  + N_o}    \Bigg)
\ea
with\\[-8mm] 
\ba
\mathcal{I}_{i,1} &= \underset{p\neq m}{\sum}\frac{P}{Md_i}\left| \mathbf{\bar{h}}_{i,i}^H \mathbf{\hat{b}}_{i,i}^{m,p} \right|^2\notag\\
\mathcal{I}_{i,2} &=\underset{k\neq i}{\sum} \overset{d_k}{\underset{p\hspace{0.5mm}=1}{\sum}}\frac{P}{Md_k}\left| \mathbf{\bar{h}}_{i,k}^H \mathbf{\hat{b}}_{i,k}^{m,p} \right|^2\notag.
\ea

Recall that in IA with perfect CSI, the conditions \eqref{SymmetricIABasicCondition3}-\eqref{SymmetricIABasicCondition2} are satisfied. Defining $\mathbf{b}_{i,k}^{m,p}   \triangleq  (\mathbf{u}_i^m)^*\circ \mathbf{v}_k^p,$ these conditions are equivalent to \\[-4mm]
 \ba
\label{SimpleEquationToBeInvoked}
  |\mathbf{ \bar{h}}_{i,i}^H\mathbf{b}_{i,i}^{m,m}|\geq c>0, \qquad\forall i,m
  \ea
  $\mathbf{ \bar{h}}_{i,i}^H\mathbf{b}_{i,i}^{m,p}=0, \forall i, \forall m\neq p,$ and $\mathbf{\bar{h}}_{i,k}^H\mathbf{b}_{i,k}^{m,p}=0, \forall k\neq i,\forall m,p,$ respectively. Naive IA entails finding vectors $\mathbf{\hat{u}}_{i}^{m}, \mathbf{\hat{v}}_{i}^{p}$ satisfying the following conditions: 
\ba
 \label{SymmetricInterferenceSISOLimitedFeedbackEquality3Modified}
|\mathbf{ \tilde{w}}_{i,i}^H\mathbf{\hat{b}}_{i,i}^{m,m}|&\geq c>0,\quad \ \forall i,m \\
\label{SymmetricInterferenceSISOLimitedFeedbackEquality1Modified}
\mathbf{ \tilde{w}}_{i,i}^H\mathbf{\hat{b}}_{i,i}^{m,p}&=0,\qquad \quad \forall i, \forall m\neq p\\
\label{SymmetricInterferenceSISOLimitedFeedbackEquality2Modified}
 \mathbf{ \tilde{w}}_{i,k}^H\mathbf{\hat{b}}_{i,k}^{m,p}&=0,\qquad \quad \forall k\neq i,\forall m, p
\ea 
where $\mathbf{ \tilde{w}}_{i,k} \triangleq [\hat{w}_{i,k}(0)\ \hat{w}_{i,k}(1)\ \ldots\ \hat{w}_{i,k}(N-1)]^T$. As noted earlier, one way to find vectors $\mathbf{\hat{u}}_{i}^{m}, \mathbf{\hat{v}}_{i}^{p}$ satisfying \eqref{SymmetricInterferenceSISOLimitedFeedbackEquality3Modified}-\eqref{SymmetricInterferenceSISOLimitedFeedbackEquality2Modified}  is provided in \cite[Appendix~III]{Cadambe2008Interference-al}. The key point here is that although based on imperfect CSI, this choice of $\mathbf{\hat{u}}_{i}^{m},\mathbf{\hat{v}}_{i}^{p},\forall i,m,p,$ results in full spatial multiplexing gain and, in addition, this can be realized with a feedback rate of $M(L-1)\log P$. We proceed with the proof of this statement.

Since $\|\mathbf{ \tilde{w}}_{i,i}\|=\|\mathbf{ \hat{w}}_{i,i}\|=1$ (from Parseval's theorem) and $\mathbf{ \tilde{w}}_{i,i},\mathbf{\hat{b}}_{i,i}^{m,p}\in\mathbb{C}^{N\times 1},\forall i,m,p$, we can always find vectors $\mathbf{q}_1,\mathbf{q}_2,\ldots,\mathbf{q}_{N-2}$ such that $\{\mathbf{ \tilde{w}}_{i,i},\mathbf{\hat{b}}_{i,i}^{m,p}/\|\mathbf{\hat{b}}_{i,i}^{m,p}\|, \mathbf{q}_1,\mathbf{q}_2, \ldots,$ $\mathbf{q}_{N-2}\}, m\neq p,$ is an orthonormal basis for $\mathbb{C}^N$. Expanding $\mathbf{\bar{h}}_{i,i}$ into this orthonormal basis, we get
 \ba
 \|\mathbf{\bar{h}}_{i,i}\|^2 &=\left|\mathbf{\bar{h}}_{i,i}^H \mathbf{ \tilde{w}}_{i,i}     \right|^2   
 +\left| \mathbf{\bar{h}}_{i,i}^H \frac{\mathbf{\hat{b}}_{i,i}^{m,p}}{\|\mathbf{\hat{b}}_{i,i}^{m,p}\|} \right|^2 +\sum_{l=1}^{N-2} \left|\mathbf{\bar{h}}_{i,i}^H\mathbf{q}_{l}\right|^2  \notag\\
 &\geq  \left|\mathbf{\bar{h}}_{i,i}^H \mathbf{ \tilde{w}}_{i,i}     \right|^2  +\left| \mathbf{\bar{h}}_{i,i}^H\frac{\mathbf{\hat{b}}_{i,i}^{m,p}}{\|\mathbf{\hat{b}}_{i,i}^{m,p}\|}  \right|^2,\quad \forall i, \forall m\neq p\notag
 \ea
which yields  
 \ba
& \frac{P}{Md_i}\left| \mathbf{\bar{h}}_{i,i}^H \mathbf{\hat{b}}_{i,i}^{m,p} \right|^2 \notag\\
&\leq  \frac{P}{Md_i}\|\mathbf{\hat{b}}_{i,i}^{m,p}\|^2 \left(\|\mathbf{\bar{h}}_{i,i}\|^2-  \left|\mathbf{\bar{h}}_{i,i}^H \mathbf{ \tilde{w}}_{i,i}     \right|^2\right)\\
 &=  \frac{P}{Md_i}\|\mathbf{\hat{b}}_{i,i}^{m,p}\|^2 \|\mathbf{\bar{h}}_{i,i}\|^2\left(1- \left|\frac{\mathbf{\bar{h}}_{i,i}^H\mathbf{ \tilde{w}}_{i,i}}{\|\mathbf{\bar{h}}_{i,i}\|}\right|^2 \right)\\
 \label{SymmetricSpatialIFAlignReplaceFreqToTime} 
  &=  \frac{P}{Md_i}\|\mathbf{\hat{b}}_{i,i}^{m,p}\|^2 \|\mathbf{\bar{h}}_{i,i}\|^2\left(1- \left|\frac{\mathbf{h}_{i,i}^H\mathbf{\hat{w}}_{i,i}}{\|\mathbf{h}_{i,i}\|}\right|^2 \right) \\ 
&\leq  \frac{P}{Md_i}\|\mathbf{\hat{b}}_{i,i}^{m,p}\|^2\left\|\mathbf{\bar{h}}_{i,i}\right\|^2(\Delta_d^{\mathrm{max}})^2\\
    \label{SymmetricSpatialIFAlignReplaceQuantizationUB} 
&\leq \frac{4P}{Md_i}\|\mathbf{\hat{b}}_{i,i}^{m,p}\|^2 \left\|\mathbf{\bar{h}}_{i,i}\right\|^2\left(\frac{1}{2^{\frac{N_d}{(L-1)}}}\right),\quad \forall i,\forall m\neq p
 \ea
 where \eqref{SymmetricSpatialIFAlignReplaceFreqToTime} follows from Parseval's theorem and \eqref{SymmetricSpatialIFAlignReplaceQuantizationUB} is obtained by invoking \eqref{MISOInterferenceFlatFadingFeedbackTempUB}.

 A similar expansion of $\mathbf{\bar{h}}_{i,k}$ into the orthonormal basis $\{\mathbf{ \tilde{w}}_{i,k},\mathbf{\hat{b}}_{i,k}^{m,p}/\|\mathbf{\hat{b}}_{i,k}^{m,p}\|, \mathbf{q}_1,\mathbf{q}_2,\ldots,\mathbf{q}_{N-2}\},k\neq i,$ yields
 \ba
\frac{P}{Md_k}\left| \mathbf{\bar{h}}_{i,k}^H \mathbf{\hat{b}}_{i,k}^{m,p} \right|^2 &\leq \frac{4P}{Md_k}\|\mathbf{\hat{b}}_{i,k}^{m,p}\|^2 \left\|\mathbf{\bar{h}}_{i,k}\right\|^2\left(\frac{1}{2^{\frac{N_d}{(L-1)}}}\right),\notag\\
\label{SymmetricSpatialIFAlignReplaceQuantizationUB2}
&\qquad\qquad\forall k\neq i,\forall m, p.
\ea
 
If we now choose $N_d=(L-1)\log P$, we get from \eqref{SymmetricSpatialIFAlignReplaceQuantizationUB} that 
\ba
\frac{P}{Md_i}\left| \mathbf{\bar{h}}_{i,i}^H \mathbf{\hat{b}}_{i,i}^{m,p} \right|^2 &\leq \frac{4P}{Md_i}\|\mathbf{\hat{b}}_{i,i}^{m,p}\|^2\left\|\bar{\mathbf{h}}_{i,i}\right\|^2\underbrace{\left(\frac{1}{2^{\frac{N_d}{(L-1)}}}\right)}_{=1/P}\notag\\
 &=\underbrace{\frac{4\|\mathbf{\hat{b}}_{i,i}^{m,p}\|^2\left\|\bar{\mathbf{h}}_{i,i}\right\|^2}{Md_i}}_{\Delta_{i,i}^{m,p}},\quad \forall i, \forall m\neq p\notag
\ea
and from \eqref{SymmetricSpatialIFAlignReplaceQuantizationUB2} that
\ba
\frac{P}{Md_k}\left| \mathbf{\bar{h}}_{i,k}^H \mathbf{\hat{b}}_{i,k}^{m,p} \right|^2 &\leq \frac{4P}{Md_k}\|\mathbf{\hat{b}}_{i,k}^{m,p}\|^2\left\|\bar{\mathbf{h}}_{i,k}\right\|^2\underbrace{\left(\frac{1}{2^{\frac{N_d}{(L-1)}}}\right)}_{=1/P}\notag\\
 &=\underbrace{\frac{4\|\mathbf{\hat{b}}_{i,k}^{m,p}\|^2\left\|\bar{\mathbf{h}}_{i,k}\right\|^2}{Md_k}}_{\Delta_{i,k}^{m,p}},\quad \forall k\neq i,\forall m,p\notag
\ea
which implies that, with $N_d=(L-1)\log P$, the overall interference power in the rate lower bound \eqref{IALimitedFeedbackSingleUserRateExpression} is upper-bounded by a constant independent of $P$ according to\\[-5mm]
\ba
\label{SymmetricSISOLimitedFeedbackInterferencePowerNew}
\mathcal{I}_{i,1}+\mathcal{I}_{i,2}& \leq   \sum_{p\neq m}\Delta_{i,i}^{m,p}+\sum_{k\neq i} \sum_{p\hspace{0.3mm}=1}^{d_k} \Delta_{i,k}^{m,p}.
 \ea
 Finally, we note that as $P\rightarrow\infty,$ the vector quantizer codebook size $2^{N_d}$ also tends to infinity and the maximum quantization error tends to zero according to \eqref{MISOInterferenceFlatFadingFeedbackTempUB}. The resolution of the vector quantizer therefore becomes arbitrarily high and we obtain $ \mathbf{\hat{w}}_{i,i}\rightarrow \mathbf{h}_{i,i} / \|\mathbf{h}_{i,i}\|,$ which implies $  \mathbf{\tilde{w}}_{i,i}\rightarrow\mathbf{\bar{h}}_{i,i}/\|\mathbf{\bar{h}}_{i,i}\|$. Substituting this into \eqref{SymmetricInterferenceSISOLimitedFeedbackEquality3Modified} yields a condition equivalent to \eqref{SimpleEquationToBeInvoked}, that is,\\[-6mm]
 \ba
  |\mathbf{\tilde{w}}_{i,i}^H\mathbf{\hat{b}}_{i,i}^{m,m}|&= \left|\frac{\mathbf{\bar{h}}_{i,i}^H\mathbf{\hat{b}}_{i,i}^{m,m}}{\|\mathbf{\bar{h}}_{i,i}\|}\right| \geq c>0,\quad \ \forall i,m \\
\Rightarrow |\mathbf{ \bar{h}}_{i,i}^H\mathbf{\hat{b}}_{i,i}^{m,m}| &\geq \|\mathbf{\bar{h}}_{i,i}\|c>0\quad \ \forall i,m.
  \ea
 Consequently, using \eqref{SymmetricSISOLimitedFeedbackInterferencePowerNew}, the spatial multiplexing gain achieved by naive IA is lower-bounded according to
\ba
&\lim_{P\rightarrow\infty}\frac{R_{\mathrm{sum}}}{\log P} \notag\\[-4mm]
&\geq \sum_{i=1}^M\overset{d_i}{\underset{m=1}{\sum}}\lim_{P\rightarrow\infty}  
 \frac{\log\left(1+  \frac{\frac{P}{Md_i}|\mathbf{\bar{h}}_{i,i}^H\mathbf{\hat{b}}_{i,i}^{m,m} |^2}{ \underset{p\neq m}{\sum}\Delta_{i,i}^{m,p} +\underset{k\neq i}{\sum} \overset{d_k}{\underset{p=1}{\sum}} \Delta_{i,k}^{m,p}  + N_o}\right)}{N\log P}\notag\\
&=\frac{\sum_i d_i}{N}\ \stackrel{t\rightarrow\infty}{\longrightarrow} \ \frac{M}{2}\qquad \text{(from \eqref{SymmetricInterferenceSISOFreqSelChoosedk} and \eqref{SymmetricInterferenceSISOFreqSelChooseN})}
\ea
which proves that full spatial multiplexing gain is achieved. We complete the proof by noting that the number of bits fed back (broadcast) by each destination for achievability of full spatial multiplexing gain using naive IA  is given, according to \eqref{IALimitedFeedbackConvertNdToNf}, by $N_f=MN_d=M(L-1)\log P$.

\bibliographystyle{IEEEtran}
\bibliography{IEEEabrv, }

\begin{thebibliography}{1}
\providecommand{\url}[1]{#1}
\csname url@samestyle\endcsname
\providecommand{\newblock}{\relax}
\providecommand{\bibinfo}[2]{#2}
\providecommand{\BIBentrySTDinterwordspacing}{\spaceskip=0pt\relax}
\providecommand{\BIBentryALTinterwordstretchfactor}{4}
\providecommand{\BIBentryALTinterwordspacing}{\spaceskip=\fontdimen2\font plus
\BIBentryALTinterwordstretchfactor\fontdimen3\font minus
  \fontdimen4\font\relax}
\providecommand{\BIBforeignlanguage}[2]{{%
\expandafter\ifx\csname l@#1\endcsname\relax
\typeout{** WARNING: IEEEtran.bst: No hyphenation pattern has been}%
\typeout{** loaded for the language `#1'. Using the pattern for}%
\typeout{** the default language instead.}%
\else
\language=\csname l@#1\endcsname
\fi
#2}}
\providecommand{\BIBdecl}{\relax}
\BIBdecl

\bibitem{Cadambe2008Interference-al}
V.~R. Cadambe and S.~A. Jafar, ``Interference alignment and the degrees of
  freedom for the {K}-user interference channel,'' \emph{IEEE Trans. Inf.
  Theory}, vol.~54, no.~8, pp. 3425--3441, Aug. 2008.

\bibitem{K.K.Mukkavilli2003On-Beamforming-}
K.~K. Mukkavilli, A.~Sabharwal, E.~Erkip, and B.~Aazhang, ``On beamforming with
  finite rate feedback in multiple-antenna systems,'' \emph{IEEE Trans. Inf.
  Theory}, vol.~49, no.~10, pp. 2562--2579, Oct. 2003.

\bibitem{Grokop2008Interference-al}
L.~Grokop, D.~N.~C. Tse, and R.~D. Yates, ``Interference alignment for
  line-of-sight channels,'' {A}vailable: http://arxiv.org/abs/0809.3035.

\bibitem{Thukral2009Spatial-multipl}
J.~Thukral and H.~B{\"o}lcskei, ``Interference alignment with limited
  feedback,'' \emph{IEEE Trans. Inf. Theory}, 2009, in preparation.

\bibitem{Love2003Grassmannian-be}
D.~Love, R.~W. {Heath Jr.}, and T.~Strohmer, ``Grassmannian beamforming for
  multiple-input multiple-output wireless systems,'' \emph{IEEE Trans. Inf.
  Theory}, vol.~49, no.~10, pp. 2735--2747, Oct. 2003.

\bibitem{Conway1996Packing-lines-p}
J.~H. Conway, R.~H. Hardin, and N.~J.~A. Sloane, ``Packing lines, planes, etc.:
  Packings in {G}rassmannian spaces,'' \emph{Exper. Math.}, vol.~5, no.~2, pp.
  139--159, 1996.

\end{thebibliography}

 \end{document}